\def\bibitemn#1#2{\bibitem{#1}#2}
\documentclass[aps,epsf,eqsecnum,feynmp,graphics,amsfonts,
latexsym,amsmath,twocolumn]{revtex4} 

\def\comment#1{}

\def\p{{\rm p}}

\def\sbf#1{\mbox{{\scriptsize${\bf #1}$}}}

\newcommand{\BF}[1]{\mbox{\boldmath $#1$}}

\begin{document}
\title{
World Nematic
Crystal Model of Gravity \\ Explaining the Absence of Torsion
 }
\author{H.~Kleinert}
 \affiliation{  Institut f\"ur Theoretische Physik.
 Freie Universit\"at Berlin, Arnimallee 14, D-14195 Berlin}
\author{J.~Zaanen}

            \affiliation{Instituut-Lorentz for Theoretical Physics,
  Leiden University, P.O.Box 9504, 2300 RA Leiden, The Netherlands }
%
%
\begin{abstract}
%
%
%
We  attribute the gravitational interaction between sources of
curvature to the world being a crystal which has undergone
 a quantum phase transition
 to a nematic phase by a condensation
 of dislocations.
  The model explains why
   spacetime has no observable torsion and predicts
the existence of curvature sources in the form of world sheets, albeit with different
high-energy properties than those of string models.

\end{abstract}

\maketitle

\section{Introduction}
Present-day string models  of elementary particles are based on
the assumption that relativistic physics will prevail at all
energy scales and, moreover, show recurrent particle spectra at
arbitrary multiples of the Planck mass. Disappointed by the
failure of these models \cite{Banks}
to explain correctly even the
the low-lying excitations, and the apparent impossibility of ever
observing the characteristic recurrences, an increasing number of
theoreticians is beginning to suspect that God may have chosen a
completely different extension of present-day Lorentz-invariant
physics to extremely high energies \cite{Volovik1,Chapline,Zhang}.
This philosophy has been advocated by one of the authors (HK) for
almost two decades. In 1987, he proposed a simple
three-dimensional euclidean
 world crystal model of gravitation
in which dislocations and disclinations
represent
 curvature and torsion in the geometry of spacetime
\cite{paper172}.
A full  theory
of gravity with torsion based on this picture
 is published
 in the
 textbook \cite{GFCM} (see also
\cite{NHOL,NHOL2}).

The simple 1987 model had the somewhat unaesthetic feature that the
crystal  possessed only  second-gradient elasticity to deliver the
correct forces between the sources of curvature, which for an
ordinary first-gradient elasticity grow linearly with the distance
$R$ and are thus confining. In this note we would like to point
out that the correct $1/R$-behavior can also be obtained in an
ordinary first-gradient world crystal with first-gradient
elasticity by assuming that the dislocations have proliferated.
This explains also why the theory of general relativity requires
only curvature for a correct description of gravitational forces,
but no torsion.  Such a state of the world crystal bears a close
relationship with the nematic quantum liquid crystals of condensed
matter physics, first suggested by Kivelson et al.
\cite{Kivelson}, and believed to be of relevance both for the
quantum Hall effect \cite{quantumhall} and in high-$T_c$
superconductors \cite{Zaanen1}. \comment{ The requirement that the
low-energy effective theory is manifestly Lorentz-invariant
implies, however, subtle differences between normal crystalline
matter and the world crystal becoming visible near the Planck
scale. Planck scale physics apparently only allows for the
topological form of nematic order discovered by Lammert, Toner,
and Rokshar\cite{Lammert}. \comment{It is important Secondly, the
notion that the world crystal breaks space-translational
invariance has to be abandoned because this would imply the
existence of massless compression modes \cite{Zaanen}.} }

Our model will be formulated as before in three euclidean
dimensions, for simplicity. The generalization to four dimensions
is  straightforward.
The elastic energy
is
expressed in terms of a material {\em displacement field\/}  $u_i({\bf x})$
as
\begin{equation}
 E =\int d^3x\, \left[ \mu
u_{ij}^2 ({\bf x})
+\frac{ \lambda }{2}u_{ii}^2  ({\bf x}) \right] ,
\label{@orig}\end{equation}
where
\begin{equation}
  u_{ij}({\bf x}) \equiv \frac{1}{2} [\partial _i u_j({\bf x}) +  \partial _j u_j({\bf x})]
\label{@}\end{equation}
 is the {\em strain tensor\/}
and
 $\mu, \nu $ are the elastic the shear moduli.
 The elastic
 energy goes to zero for infinite wave length since in this
 limit $ u_i({\bf x})$ reduces to a pure translation under which
the
 energy of the system is invariant. The
crystallization process
causes a spontaneous breakdown of the
translational symmetry of the system.
The elastic distortions describe the Nambu-Goldstone modes
resulting from this symmetry
breakdown.
Note that so far the crystal has an extra
longitudinal sound wave with a different  velocity than the shear waves.

A crystalline material always contains defects. In their presence,
the elastic energy is
\begin{equation}
  E = \int d^3 x \left[
\mu (u_{ij} - u^{\rm p}_{ij})^2+
\frac{ \lambda }{2} (u_{ii} - u^{\rm p}_{ii})^2\right],
\label{@elen}\end{equation}
 where $u_{ij}^\p$ is the so-called {\em plastic strain
tensor\/}  describing the defects. It is composed of an ensemble
of
 lines with a dislocation density
\begin{equation}
    \alpha _{il} = \epsilon _{ijk }
      \partial _j \partial _k u_l({\bf x}) =  \delta _i
     ({\bf x};L) (b_l +  \epsilon _{lqr}  \Omega _q x_r).
\label{DLD}\end{equation}
and a {\em disclination density\/}
\begin{equation}
 \theta _{il} =  \epsilon _{ijk} \partial _j \phi^{\rm p}_{kl}
   =  \delta _i ({\bf x};L)  \Omega _l,
\label{DED}\end{equation}
where $b_l$ and $\Omega_l$ are the so-called Burgers and Franck
vectors of the defects. The densities satisfy the conservation
laws
\begin{equation}
\partial _i  \alpha _{ik} = -  \epsilon _{kmn}
    \theta _{mn},
    ~~~
      \partial _i   \theta _{il} = 0.
\label{DED}\end{equation}
 Dislocation lines are either closed or they end in disclination
 lines, and
 disclination lines are closed.
These are Bianchi identities
 of the defect
 system.

An important geometric quantity characterizing dislocation
 and disclination lines is the {\em  incompatibility\/}
 or {\em defect density\/}
\begin{equation}
  \eta_{ij} ({\bf x}) =  \epsilon _{ikl}  \epsilon _{jmn} \partial _k
      \partial _m u_{ln}^P({\bf x}).
\label{@inc}\end{equation}
It can be decomposed into disclination and dislocation density as
follows \cite{GFCM}:
\begin{equation}
\eta _{ij}({\bf x}) \!= \! \theta _{ij}({\bf x})\!+ \!
\frac{1}{2} \partial _m \left[
      \epsilon _{min}  \alpha _{jn}({\bf x})\! +\! (i\!\leftrightarrow\! j) \!- \! \epsilon _{ijn}
 \alpha _{mn}({\bf x})\right].
\label{@etaten}\end{equation}

 This tensor is symmetric and conserved
\begin{equation}
 \partial _i \eta_{ij}({\bf x}) = 0,
\label{@}\end{equation}
again a Bianchi identity of the defect
 system.

It is useful to separate from the
dislocation density
(\ref{DLD})
the contribution from the disclinations
which causes the nonzero right-hand side of
(\ref{DED}).
Thus we
 define
a {\em pure
dislocation density\/}
\begin{equation}
 \alpha^{b}_{ij}({\bf x})\equiv
 \alpha_{ij}({\bf x})-
 \alpha^{\Omega}_{ij}({\bf x})
\label{@}\end{equation}
which satisfies
 $\partial _i\alpha^{b}  _{ij}=0$.
Accordingly, we split
\begin{equation}
 \eta _{ij}({\bf x})=
 \eta _{ij}^{b}({\bf x})+
 \eta _{ij}^{\Omega}({\bf x}),
\label{@}\end{equation}
where
\begin{equation}  \!\phantom{x}
 \eta _{ij}^{b}({\bf x})\!=\!
\frac{1}{2}\left[
      \epsilon _{min}  \alpha _{jn}^{b}({\bf x}) + (i\leftrightarrow j) -  \epsilon _{ijn}
 \alpha _{mn}^{b}({\bf x})\right]\!, ~\!\!\!\!\!\!
\label{@}\end{equation}
and the pure disclination part of the defect tensor
looks like
(\ref{@etaten}), but with superscripts
$ \Omega $ on
$ \eta _{ij}$ and $ \alpha _{ij}$.

The tensors $ \alpha _{ij},\,  \theta _{ij }$, and $\eta_{ij}$ are
linearized versions of important geometric tensors in the {\em
Riemann-Cartan space\/} of defects, a noneuclidean space
  with
curvature and torsion.
Such a space can be generated from a flat space
by a plastic distortion,
which is mathematically
represented by a
{\em nonholonomic\/} mapping \cite{NHOL,NHOL2}
%
$  x_i \rightarrow  x_i + u_i({\bf x}).$
%
Such a mapping is nonintegrable.
The displacement fields and their first derivatives
fail to satisfy the Schwarz integrability criterion:
\begin{eqnarray}  \!\!\!\!\!\!
\left(\partial _i \partial _j \!-\! \partial _j \partial _i\right)
     u({\bf x})   \neq  0 ,~~
\left( \partial _i \partial _j \!-\! \partial _j \partial _i\right)
    \partial _k u_l ({\bf x})  \neq  0.
\label{@}\end{eqnarray}
The
metric
and
the affine connection
of the geometry in the plastically distorted space are
$g_{ij} = \delta _{ij}+ \partial _i u_j + \partial _j u_i
$
and
%
$ \Gamma _{ijl} = \partial _i \partial _j u_l,$
%
respectively.
The noncommutativity of the derivatives in front
 of $u_l({\bf x})$ implies a nonzero torsion,
the torsion tensor being
%
$S_{ijk}\equiv ( \Gamma _{ijk}- \Gamma _{jik})/2.$
%
The dislocation density
$ \alpha _{ij}$ is
equal to
%
$  \alpha _{ij}=\epsilon _{ikl} S_{klj},$
%

The noncommutativity of the derivatives in front
 of $\partial _k u_l({\bf x})$ implies a nonzero
curvature.
The disclination density $ \theta _{ij}$ is the Einstein
 tensor
%
$\theta _{ij}=
R_{ji} -\frac{1}{2} g_{ji}R
$ 
of this Einstein-Cartan
defect geometry.
The tensor $\eta_{ij}$, finally, is the
Belinfante symmetric  energy momentum tensor,
 which is defined in terms of the canonical
 energy-momentum
tensor  and the spin density by a relation just like
(\ref{@etaten}).
    For more details on the
geometric aspects
see
 Part IV in Vol. II of
\cite{GFCM},
where the full one-to-one
correspondence between
defect systems and Riemann-Cartan geometry is developed as well as
a
gravitational
theory
 based on this analogy.

Let us now show how linearized gravity emerges
from the energy
(\ref{@elen}).
For this we
eliminate the jumping
surfaces
in the defect gauge fields
from the partition function by introducing
 conjugate variables and
associated stress gauge fields. This is done by
rewriting the elastic action of defect lines as
\begin{eqnarray}
 \!\!\!\!\!\!\!\!\! E \!=\! \int d^3 x \left[\! \frac{1}{4\mu} \!
\left(\sigma _{ij}^2\!-\!\frac{ \nu }{1+ \nu }
\sigma _{ii}^2\right)
\!+\! i \sigma _{ij}
(u_{ij}
\!-\!u_{ij}^P)
\right]
,
\label{@EN0}\end{eqnarray}
where $ \nu \equiv  \lambda /2( \lambda +\mu)$
is Poisson's ratio,
and forming the partition function,
integrating
the Boltzmann factor $e^{-E/k_BT}$
 over $\sigma_{ij},  u_i$,
and summing over all jumping surfaces $S$
in the plastic fields.
The integrals
over  $u_i$ yield
the conservation law
%
$   \partial _i  \sigma  _{ij} = 0.$
%
This can
 be enforced as a Bianchi identity
 by
 introducing a stress gauge field
$h_{ij}$ and writing
%
$  \sigma _{ij} = G_{ij}\equiv
\epsilon _{ikl}\epsilon _{jmn}
\partial _k
\partial _m h_{ln}.$
%
The double curl on the right-hand side
 is
recognized as
 the Einstein tensor
in the geometric description of
stresses, expressed in terms of
a small deviation n $h_{ij}\equiv
g_{ij}- \delta _{ij}$
of the metric
from the
flat-space form.
Inserting $G_{ij}$
into (\ref{@EN0})
and using
(\ref{@inc}),
we can
replace the energy in the partition function
by
$
 E =E^{\rm stress}
 +E^{\rm def}$ where
\begin{equation}
E^{\rm stress}
 \!+\!E^{\rm def}
\! \equiv \!
 \int d^3 x \left[ \frac{1}{4\mu}
\left(G _{ij}^2\!-\!\frac{ \nu }{1+ \nu }
G _{ii}^2\right)
 +i
h _{ij} \eta _{ij}
\right],
\label{@ENN}\end{equation}
where the defect tensor
 (\ref{@etaten})
has the decomposition
\begin{equation}\!\!\!\!\!\!
 \eta _{ij}
=\eta^ \Omega _{ij}+
\partial _m
      \epsilon _{min}  \alpha ^b_{jn}
.
\label{@En3}\end{equation}
The defects have also  core energies which
has been ignored so far.
Adding these
for the dislocations
and ignoring, for a moment, the disclination part of the defect density
in (\ref{@En3}),
 we obtain
\begin{equation}
 E^{\rm disl}\!=\!i \int d^3 x \left(
      \epsilon _{imn}\partial _m
h _{ij}
 \alpha ^b_{jn}+\frac{ \epsilon _c}2{\alpha _{jn}^{b\,2}}
\right)
.      ~ \!\!\!\!
\label{@EN3}\end{equation}
We now assume
that the world crystal has undergone a transition
to a condensed phase in which dislocations are condensed.
To reach such a state, whose existence was
conjectured for two-dimensional crystals
in Ref.~\cite{HN}, the model
requires a modification by an additional
rotational energy, as shown in \cite{RST}
and verified by Monte Carlo simulations in
\cite{RST2}.
The three-dimensional extension of the
model is described in \cite{GFCM}.

The condensed phase is described by a
partition function
in which
the discrete sum over the pure dislocation densities
in $\alpha ^b_{jn}$ is
approximated by
an ordinary functional integral.
This has been shown
in Ref.~\cite{NHOL}.
The general integration rule is
\begin{equation}
\int d^3l\, \delta (\BF{\partial}\cdot {\bf l})\,e^{- \beta {\sbf l}^2/2+i  {\sbf l}
{\sbf a}}=e^{-{\sbf a}_T^2/2 \beta },
\label{@intrule}
\end{equation}
where
${\bf a}_T$
has the components
$a_{Ti}\equiv   -
i \epsilon _{ijk}\partial _ja_k/ \sqrt{-{\BF \partial}^2}$.
The Boltzmann
factor  resulting in this way from
                        $E^{\rm stress}$ plus
(\ref{@EN3})
has now
the energy
\begin{equation}
 E' = \int d^3 x \left[ \frac{1}{4\mu}
\left(G _{ij}^2\!-\!\frac{ \nu }{1+ \nu }
G _{ii}^2\right)
 +
\frac{1}{2 \epsilon _c}
G _{ij}\frac{1}{-{\BF \partial}^2}
G _{ij}
\right]
.
\label{@EN31}\end{equation}
The second term implies  a Meissner-like screening
of the initially confining gravitational forces between
the disclination part of the defect tensor
to Newton-like forces.
For distances longer than the Planck
scale,
we may
ignore the stress term
and find the effective gravitational action
for the disclination part of the defect tensor:
\begin{equation}
 E \approx
\int d^3 x \left(
\frac{1}{2 \epsilon _c}
G _{ij}\frac{1}{-{\BF \partial}^2}
G _{ij}+i h_{ij}\eta^ \Omega _{ij}
\right)
.
\label{@EN32}\end{equation}
 A path integral
over $h_{ij}$ and a sum over all line ensembles applied to the
Boltzmann factor $e^{-E/\hbar}$ is a simple Euclidean model of
pure quantum gravity. The line fluctuations of $\eta^ \Omega
_{ij}$ describe a fluctuating Riemann geometry perforated by a
grand-canonical ensemble arbitrarily shaped lines of curvature. As
long as the loops are small they merely renormalize the first term
in the energy (\ref{@EN32}). Such effects were calculates in
closely related theories in great detail in Ref.~\cite{KM}. They
also give rise to post-Newtonian terms in the above linearized
description of the Riemann space.

We may now add matter to this gravitational environment. It is
coupled by the usual Einstein interaction
\begin{equation}
 E^{\rm int} \approx
\int d^3 x \, h_{ij} T^{ij} , \label{@EN4}\end{equation} where  $
T ^{ij} $ is the symmetric Belinfante energy momentum tensor of
matter. Inserting for $G_{ij}$ the double-curl of $h_{ij}$ we see
that the energy (\ref{@EN32}) produces the correct Newton law if
the core energy is $ \epsilon _c=8\pi G$, where $G$ is Newton's
constant.

Note that the condensation process of dislocations has led to a
pure Riemann space without torsion. Just as a molten crystal shows
residues of the original crystal structure only at molecular
distances, remnants of the initial
 torsion could be observed only near the Planck scale.
This explains
why present-day
 general relativity
requires only a Riemann space, not a Riemann-Cartan space.

 In the non-relativistic context, a
dislocation condensate is characteristic for a nematic liquid
crystal, whose order is translationally invariant, but breaks
rotational symmetry (see \cite{HN,GFCM} in the two dimensions and
\cite{Zaanen} in the 2+1-dimensional quantum theory). The Burgers
vector of a dislocation is a vectorial topological charge, and
nematic order may be viewed as an ordering of the Burgers vectors
in the dislocation condensate. Such a manifest nematic order would
break the low energy Lorentz-invariance of space-time. We may,
however, imagine that the stiffness of the directional field of
Burgers vectors is so low that, by the criterion of
Ref.~\cite{KleiCrit}, they have undergone a Heisenberg-type of
phase transition into a directionally disordered phase
 in an environment with only a few disclinations.
In three dimensions, dislocations (and disclinations) are
line-like. This has the pleasant consequence, that they can  be
described by the disorder field theories developed in \cite{rem2}
in which the proliferation of disclinations follows the typical
Ginzburg-Landau pattern of the field expectation acquiring a
nonzero expectation value. A cubic interaction becomes isotropic
in the continuum limit \cite{rem3} (this is the famous
fluctuation-induced symmetry restoration of the Heisenberg fixed
point in a $\phi^4$-theory with O(3)-symmetric plus cubic
interactions \cite{KS}. The isotropic phase is similar to what has
been called a {\em topological} form  of nematic order by Lammert
et al. \cite{Lammert} in a generalized $Z_2$ gauge theory of
nematic order. Also there rotational (Lorentz-) invariance is
restored even though there is no condensate of disclinations, and
the Burgers vectors are disordered by fluctuations (see also
ref.~\cite{Zaanen}). The dual of this phase is Coulomb-like.

The above description of defects was formulated in what has been
named tangential approximation to the Euclidean group \cite{GFCM},
in which the discrete rotations are treated as if they took place
in the tangential place with arbitrary real Franck vectors
$\Omega^i$ [see Eq.~(\ref{DLD})]. In a more accurate formulation,
the nonabelian nature of the rotations and the quantization of
$\Omega_i$ must be taken into account. Their discreteness is
certainly remembered in the nematic phase, even if the directions
of the Burgers vectors become diordered (see also the discussion
in Ref.~\cite{Bais}). This implies that there are elements of
quantized curvature fluctuating in spacetime. Fortunately, this
does not introcduce any unphasical results at  presently
accessible length scales since
 these
fluctuations mainly renormalize the basic curvature energy in
(\ref{@EN32}), as discussed before.

\comment{In (\ref{@EN31}), all components of the stress tensor
$G_{ij}$ have acquired the same mass which makes both shear- and
compression stresses short-ranged. This explains why none of the
massless modes of the initial world crystal (\ref{@orig}) is
observed in the low-energy universe. This is in contrast to
non-relativistic crystalline matter, where shear rigidity is
associated with translational symmetry breaking, and the
dislocation condensate gives a Meissner-Higgs mass only to shear
modes. A non-relativistic dislocation fluid is a superfluid
possessing a residual massless compression mode \cite{Zaanen}. The
reason for the decoupling of the dislocation currents from the
compression sector of the stress gauge fields is a consequence of
the non-relativistic glide constraint which expresses the fact
that dislocations only propagate in the directions of their
spatial Burgers vector. Orthogonal directions would involve an
excessive migration of atoms in the crystal. This constraint is
absent in  pure Einstein gravity, which differs in this respect
essentially from a world-crystal made up by material points which
would have a massless compression mode. If it turns out that a
full understanding of gravitational forces does require an extra
massless scalar field as suggested first by Brans and Dicke
\cite{BD}, and recently in the form of {\em quintessence\/} to
explain the apparently accelerated expansion of the present
universe, this would be a definite signal towards a material
nature of the world crystal with glide condition. }

The generalization to four Euclidean spacetime dimensions changes
mainly the geometry of the defects.
 In
four dimensions, they become world sheets, and a second-quantized
disorder field description of surfaces has not yet been found. But
the approximation of representing a sum over dislocation surfaces
in the proliferated phase as an integral as in
Eq.~(\ref{@intrule}) will remain valid, so that the above line of
arguments will survive, this being a natural generalization of the
Meissner-Higgs mechanism. The disclination
sources of curvature will
be world sheets, as an attractive feature for string theorists.
However, the high-energy properties will be completely different.
On the one hand,
 these surfaces behave nonrelativistically
as the energies approach the Planck scale,
on the other hand
they will not have the characteristic multi-Planck recurrences
of the common strings.
Although the latter property may never be verified in the laboratory,
the deviations from relativity at high energies
or short distances
may come into experimentalists reach
in the not too distant future.

Note that our model has automatically a vanishing cosmological constant.
Since the atoms in the crystal are in equilibrium, the pressure
is zero. This explanation is similar to that given by Volovik \cite{Volovik1}
with his helium droplet  analogies.

%


\end{document}